\documentclass{aa}  
\def\ha{H$\alpha$}

\def\hb{H$\beta$}
\def\oiii{[O {\sc iii}]}
\def\oi{[O {\sc i}]}
\def\nii{[N\,{\sc ii}]}
\def\sii{[S\,{\sc ii}]}
\def\feii{[Fe\,{\sc ii}]}

\def\hh{H$_2$}
\def\m29{\object{M~2--9}}
\def\magcir{\ \raise-2.truept\hbox{\rlap{\hbox{$\sim$}}\raise5.truept
    \hbox{$>$}\ }}
\def\kms{\relax \ifmmode {\,\rm km\,s}^{-1}\else \,km\,s$^{-1}$\fi}
\def\deg{$^\circ$}
\usepackage{graphicx}
\usepackage{longtable,lscape}
\usepackage{txfonts}
\usepackage{hyperref}
\begin{document}
\title{The evolution of M~2--9 from 2000 to 2010
\thanks{Based on observations obtained at; the 2.6 Nordic
    Optical Telescope operated by NOTSA; the 2.5m~INT and 4.2m~WHT
    telescopes of the Isaac Newton Group of Telescopes in the Spanish
    Observatorio del Roque de Los Muchachos of the Instituto de
    Astrof\'\i sica de Canarias. 
Also based on observations made with the
NASA/ESA Hubble Space Telescope, obtained at the Space Telescope
Science Institute, which is operated by the Association of
Universities for Research in Astronomy, Inc., under NASA contract
NAS 5-26555. These observations are associated with programs 8773
and 9050.
}\thanks{Figure 5b is only available in electronic form via
http://www.edpsciences.org}}

\author{Romano L.M. Corradi\inst{1,2}
          \and
        Bruce Balick\inst{3}
           \and
        Miguel Santander--Garc\'\i a\inst{4,1,2}
	}

   \offprints{R. Corradi}

   \institute{
Instituto de Astrof{\'{\i}}sica de Canarias, 38200 La Laguna, 
Tenerife, Spain \email{rcorradi@iac.es}
   \and
Departamento de Astrof\'\i sica, Universidad de La Laguna, 
38206 La Laguna, Tenerife, Spain 
   \and
Astronomy Department, University of Washington, Seattle, WA 98195, USA
   \and
Isaac Newton Group of Telescopes, Apart. de Correos 321, 38700
Santa Cruz de la Palma, Spain
             }

\date{Received 20-12-2010 / Accepted 10-02-2011}

\abstract
{Understanding the formation of collimated outflows is one of the most
  debated and controversial topics in the study of the late stages of
  stellar evolution.}
{\m29\ is an outstanding 
  representative of extreme aspherical flows. It presents  
  unique features such as a pair of high-velocity dusty polar blobs and a 
mirror-symmetric rotating  pattern in the inner lobes. 
Their study provides important information on the nature of the
  poorly understood central source of \m29\ and its nebula.}
{Imaging monitoring at sub-arcsec resolution of the evolution of the 
nebula in the past decade is presented. 
Spectroscopic data provide complementary information.}
{We determine the proper motions of the dusty blobs, which infer a 
new distance estimate of 1.3$\pm$0.2~kpc, a total nebular 
size of 0.8~pc, a speed of 147~\kms, and a kinematical age of 2500~yr. 
The corkscrew geometry of the inner rotating pattern is confirmed and 
quantified. Different recombination timescales for different ions explain 
the observed surface brightness distribution. 
According to  the images taken after 1999, the pattern rotates with 
a period of $92\pm4$~years. On the other hand, the analysis of images taken
between 1952 and 1977 measures a faster angular velocity.
If the phenomenon were related to orbital motion, 
this would correspond to a modest orbital eccentricity ($e=0.10\pm0.05$), 
and a slightly shorter period ($86\pm5$ years).
New features have appeared after 2005 on the west side 
of the lobes and at the base of the pattern.} 
{The geometry and travelling times of the rotating pattern support our 
previous proposal that the phenomenon is produced by a collimated spray of 
high velocity particles (jet) from the central source, which excites the 
walls of the inner cavity of \m29, rather than by a ionizing photon beam. 
The speed of such a jet would be remarkable: between 11\,000 and 16\,000~\kms. 
The rotating-jet scenario may explain the formation and excitation of most 
of the features observed in the inner nebula, with no need for additional 
mechanisms, winds, or ionization sources.
All properties point to a symbiotic-like interacting binary as the central 
source of \m29. The new distance determination implies 
system parameters that are consistent with this hypothesis.}

\titlerunning{the evolution of M~2--9 from 2000 to 2010}
\authorrunning{Corradi, Balick \& Santander-Garc\'\i a}
\maketitle


\begin{figure}
\centering
\includegraphics[width=6.7cm]{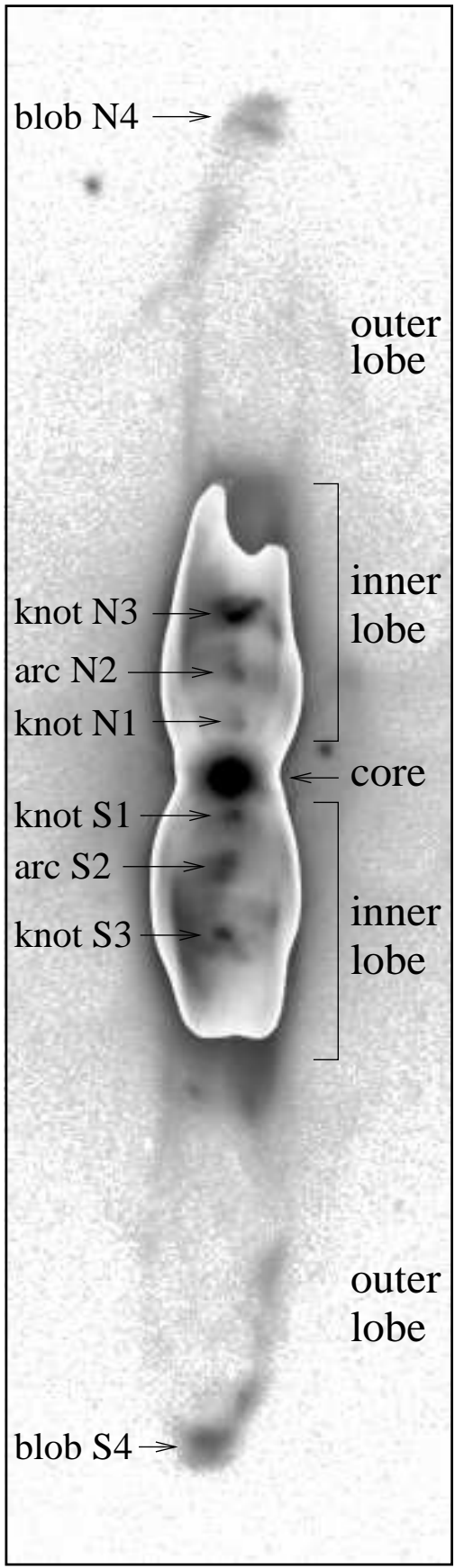}
\caption{Composition of \ha+\nii\ images of \m29\ with the
  identification of features discussed throughout this paper. Where
  possible, we have tried to maintain the nomenclature used by
  previous authors. North is up and east is left. The 
field of view (f.o.v.)  is 40$''$$\times$140$''$.}
\label{F-cartoon}
\end{figure}

\section{Introduction}

\m29, also known as the ``Butterfly'' or the ``Twin Jet'' nebula, is a
bright bipolar nebula of uncertain nature. It is included in the
general list of Galactic planetary nebulae as PN G010.8+18.0
(\cite{a92}), and is often considered to be a young planetary nebula.
However, the spectrum of its nebular nucleus (\cite{b89}) and
  the rotating pattern in the inner lobes (e.g. \cite{ls01}) made
  several authors suggest that \m29\ contains a symbiotic star at its
  centre. \cite{sk01} supported this classification based on
  infrared colours typical of symbiotic Miras; however, several
  authors have pointed out that these colours can also be produced by
  dusty discs (\cite{ch07, co11, l11}).

Figure~\ref{F-cartoon} illustrates the morphology of \m29.  The
  image is a composition of short and deep \ha+\nii\ exposures (see
  Table~\ref{T-log}): the 2009 image from the NOT telescope is used to
  display the inner regions of the nebula (inset picture), and the
  outer faint regions are from the 2006 INT image.  The object is
extraordinary in several aspects. At large distances from the nucleus
along the main axis of the nebula, faint outer lobes with a 
point-symmetric surface-brightness distribution are observed. They
culminate in high--velocity dust blobs that scatter and absorb the
illuminating light like a pair of fog clouds receding in opposite
directions from a streetlight, as shown by \cite{s97}. A 
  space velocity of the dusty blobs of 164~\kms, a total separation of
  0.4~pc, and an age of 1200~yr were determined by \cite{s97} for the
  adopted distance of 650~pc (twice as short as derived in this
  paper).
But the outstanding characteristic of \m29\ is undoubtedly its moving
pattern in the inner lobes (features S1, S2, N1 and N2), rotating like
in a lighthouse with a period of about a century (Aller and Swings
1972, Kohoutek \& Surdej 1980, Doyle et al. 2000).  No other similar
objects are known in the sky, with the possible exception of
$\eta$~Carinae, where azimuthal illumination variability might also
occur (\cite{sm04}). However, in $\eta$~Car the phenomenon is far less
clear and regular than in \m29.  

The nebula of \m29\ is oriented in the sky at position angle
358\deg\ and is seen close to edge-on: an inclination of its symmetry
axis on the plane of sky of 11\deg\ (Goodrich 1991), 15\deg\ (Schwarz
et al. 1997), and 17\deg\ (Solf 2000) was determined. A detailed study
of the kinematics of the nebula is presented in \cite{s00}.

In this article, we present new images monitoring the rotation of the
pattern in the past decade. They also allow us to measure the
separation rate of the outer blobs, which in turn allows the distance
to \m29\ to be determined via a straightforward geometrical
method. This information provides new constraints on the origin of the
phenomena observed in \m29.

\section{Observations}

\subsection{Images}

The images of \m29\ obtained before 2000 are presented and discussed
in \cite{d00}. Since then, imaging monitoring at sub-arcsec resolution
has continued mostly at the 2.6m Nordic Optical Telescope (NOT) on La
Palma, with a deep image also being taken at the 2.5m Isaac Newton Telescope
(INT).  At the NOT, the ALFOSC instrument was used, providing a
spatial scale of 0$''$.19 per pixel.  At the INT, we used the prime
focus Wide Field Camera, which has 0$''$.33 pixels.

HST WFPC2 \oiii\ and \nii\ images of the nebula were obtained on June
2001 (proposal ID 8773) and September 2001 (proposal ID 9050). The
latter ones were taken with the nebula centred on the PC1 detector,
using a four-point dithering sequence and by carefully avoiding
saturation of the core of \m29, allowing the maximum spatial
resolution of HST to be obtained for the study of the innermost
regions of the nebula.

Details of the observations, including the central wavelength
$\lambda_c$ and full width at half maximum (FWHM) of the filters used,
exposure times and seeing, are given in Tab.~\ref{T-log}.  We note that at
both the NOT and the INT the \ha\ images include emission from the
nearby \nii\ doublet, whose contribution is significant throughout the
nebula.

\begin{table}[!ht]
\caption{Log of the new images of \m29. See Tab.~1 in \cite{d00} for
  our previous imagery.}
\begin{tabular}{cccccc}
\hline\hline
Date    & Tel. & Filter  & $\lambda_c$/FWHM & Exp. time  & Seeing \\
        &      &     &                   [\AA]  &  [sec]    & [$''$] \\
\hline\\[-7pt]                                                    
2000.47 & NOT & \ha+\nii & 6571/47  &      180   & 0.9  \\
2002.49 & NOT & \ha+\nii & 6571/47  &  60,1200   & 0.7  \\
2006.38 & INT & \ha+\nii & 6568/95  &       5400 & 1.0  \\
2006.43 & NOT & \ha+\nii & 6577/180 & 30,240,900 & 0.8  \\
2007.26 & NOT & \ha+\nii & 6577/180 & 10,30,360  & 0.8  \\
2008.48 & NOT & \ha+\nii & 6577/180 & 540        & 1.0  \\
2009.28 & NOT & \ha+\nii & 6577/180 & 10,540     & 0.6  \\
2010.42 & NOT & \ha+\nii & 6577/180 & 10,540     & 0.6  \\[4pt]
2000.47 & NOT & \oiii    & 5012/30  &   240,600  & 1.0  \\
2001.49 & HST & \oiii    & 5012/27  & 600        & --   \\
2001.73 & HST & \oiii    & 5012/27  & 1040       & --   \\
2002.49 & NOT & \oiii    & 5012/30  &   30,300   & 0.9  \\  
2006.43 & NOT & \oiii    & 5010/43  &   30,900   & 0.8  \\
2007.26 & NOT & \oiii    & 5010/43  &  60,1500   & 0.9  \\
2008.48 & NOT & \oiii    & 5010/43  &  1200      & 0.9  \\
2009.28 & NOT & \oiii    & 5010/43  &  1500      & 0.6  \\
2010.42 & NOT & \oiii    & 5010/43  &  1500      & 0.8  \\[4pt]
2001.49 & HST & \nii     & 6591/29  & 1300        & --   \\
2001.73 & HST & \nii     & 6591/29  &  400        & --   \\
\hline
\end{tabular}
\label{T-log}
\end{table}

\subsection{Spectra}

Deep medium-resolution spectra were obtained on April 8, 2006, at the
4.2m William Herschel Telescope (WHT) on La Palma, using the
double-arm long-slit spectrograph ISIS. The slit was positioned along
P.A.=$-1$\deg\  to cover the main knots of the rotating pattern in
the inner lobes, as well as the outermost dusty blobs. Slit width was
0$''$.6.  In the ISIS red arm, the R1200R grating was used, providing a
spectral coverage from 6330~\AA\ to 7360~\AA, a reciprocal dispersion of
0.22~\AA\ pix$^{-1}$, and a resolution of 28~\kms.  In the blue arm,
the selected grating R1200B provided a spectral coverage between 4465~\AA\ 
and 5340~\AA, a reciprocal dispersion of 0.22~\AA\ pix$^{-1}$, and a
resolution of 32~\kms. Several red and blue spectra were obtained
simultaneously using the standard dichroic of ISIS, with exposure
times of 30 sec, 400 sec $\times$ 2, and 1800 sec $\times$3. Seeing was
1$''$.2 FWHM.

\section{Analysis of the motion of the dusty blobs}
\label{S-blobs}


\cite{s97} presented the analysis of the outermost blobs of \m29 (S4 and
N4 in Figure~\ref{F-cartoon}).  These radially opposed features are
60\%\ linearly polarized and both are redshifted with respect to the core,
indicating that they are mainly composed of dust and that the
radiation from them is reflected light rather than intrinsic emission.
The proper motion analysis of the blobs by \cite{s97} was based on
images taken in the period from 1978 and 1994. While the baseline was
significant (16 years), the astrometric calibration was rather
uncertain, especially for the first epoch for which the separation of
the blobs could not be measured, and was taken as quoted in the
original paper by \cite{ks80}. The new deep \ha\ images obtained on
2002 and 2006 allowed a more precise analysis to be performed. All CCD
images in our hands, spanning 18.1 years in the period from 1988 to
2006, were first registered astrometrically using the USNO-B1.0
catalogue (\cite{m03}), and then rescaled to a common orthogonal grid
with pixel size of 0$''$.36. In this reference frame, the position of
the central source and two nearby reference field stars in the
different images show a maximum variation of 0$''$.1 in right
ascension and 0$''$.2 in declination. These are conservatively taken
as the uncertainties in the position of the blobs and the central
star. The apparent separation of the blobs was then computed in two
ways.

\begin{figure}[!ht]
\centering
\includegraphics[width=9cm]{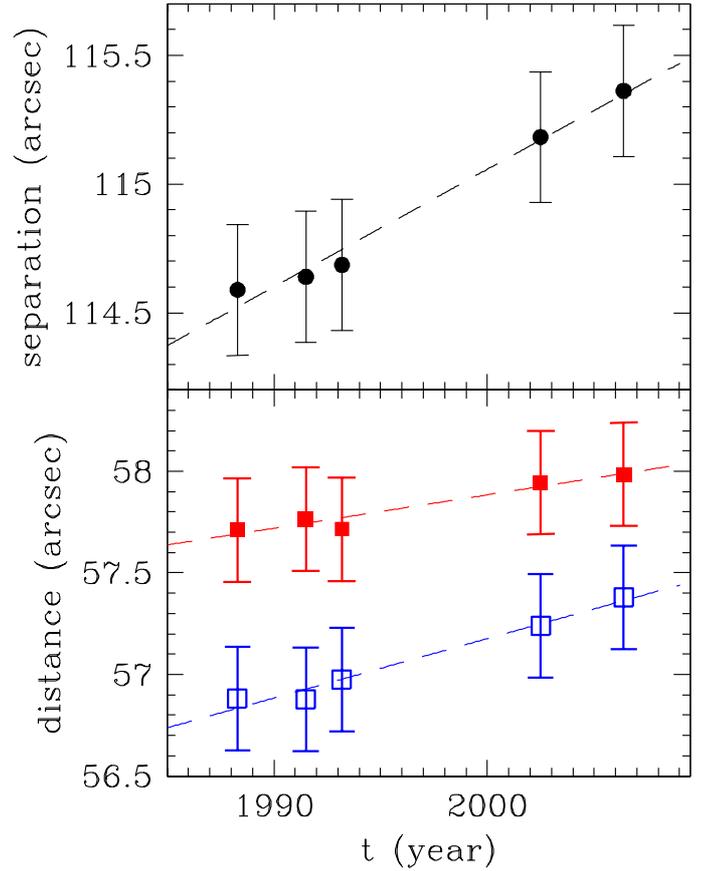}
\caption{Proper motions of the dusty blobs of \m29. Top: the total
  interblob separation. Bottom: the distance from the central
  source for each blob separately. Core separation for the north and south
  blobs is indicated by the filled and empty squares,
  respectively. Least squares fits are shown by the dashed lines.
  Positions are computed using the intensity-weighted barycentre.}
\label{F-propermotion}
\end{figure}

\subsection{First moment analysis}

Our first step was to compute the barycentre of the emission for each
blob, defined as the intensity-weighted first moments in a box of 
fixed size centred on the blob.  The total separation of the blobs
(Figure~\ref{F-propermotion}, top) is well fitted by a straight line
with slope of $0.046\pm0.004$ arcsec~yr$^{-1}$.  This is similar to
the determination by \cite{s97} of $0.051\pm0.007$ arcsec~yr$^{-1}$.

If we consider the motion of each blob separately and measure their
separation from the central star (Figure~\ref{F-propermotion},
bottom), a difference seems to emerge with the southern blob moving
faster, at a rate of $0.029\pm0.003$ arcsec~yr$^{-1}$ compared to the
$0.016\pm0.003$ arcsec~yr$^{-1}$ of the northern one. The difference
is however within $\sim$2$\sigma$ of the errors which, as
  discussed above, also include the uncertainty associated with the
image registration process.  We also note in Figure~\ref{F-propermotion}
that the southern blob is more than half an arcsec closer to the
central star than the northern one.

\subsection{Magnification method}

The separation rate of the blobs was also determined by means of the
so-called magnification method (see e.g. \cite{bh04}). Basically, the
first-epoch image is rescaled using a magnifying factor $M$ until the
residuals of the subtraction from a more recent image are
minimized. This magnification factor was computed between each pair of
subsequent images, as well as by comparing the first-epoch and
last-epoch images.  The result is that the magnification in the 18.08
years considered is $M_N=1.0046$ for the northern blob (corresponding
to a proper motion of $0.015$ arcsec~yr$^{-1}$), and $M_S=1.0088$ for
the southern one ($0.028$ arcsec~yr$^{-1}$). The uncertainty is of the
order of 0.0004 in both cases. These figures fully agree with the
separation rates computed from the first moment analysis.
The lower accuracy of the latter method is mainly due to the 
lack of a clear separation of the blobs from the outer lobes. 
Combined with the variable seeing, this prevents homogeneous 
definition of the blob boundaries in the different images.

\begin{figure}[!ht]
\centering
\includegraphics[width=9cm]{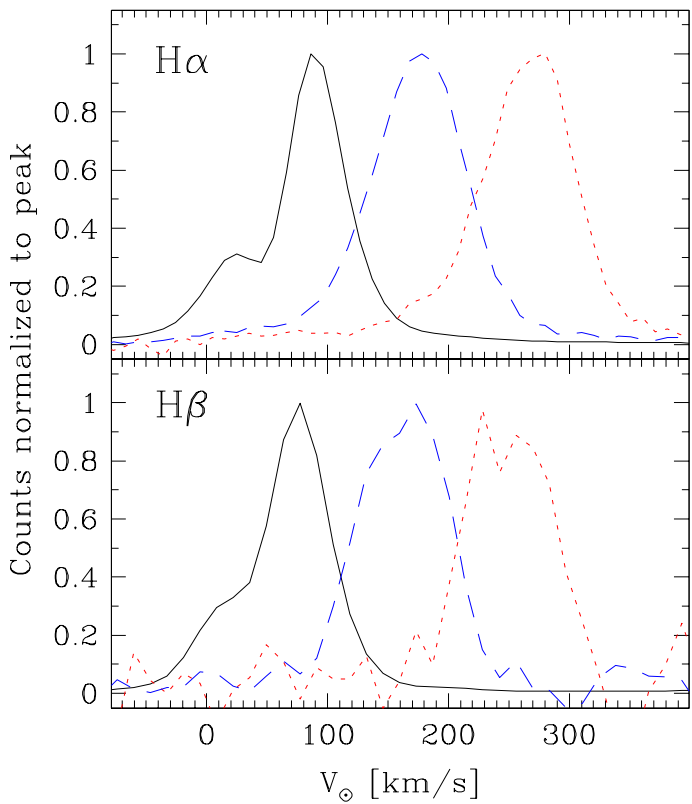}
\caption{\ha\ (top) and \hb\ (bottom) line profiles for the core of
  \m29\ (black solid line), north blob (red dotted line), and south
  blob (blue dashed lines). Velocities in the abscissae 
are heliocentric. All profiles have been
  normalized for easier comparison. Note that local structures in the
  \hb\ profiles of the blobs are not real, but caused by a modest signal-to-noise
  ratio.}
\label{F-lineprofile}
\end{figure}

\subsection{Line-of sight velocities}

Remembering that the emission of the outermost blobs is 
scattered light from the central source (Schwarz et al. 1997), their 
observed redshift is the sum of the velocity of recession of each blob
with respect to the central object, $V_{exp}$, plus the component of
this velocity along the line of sight (see figure 5 in Schwarz et
al. 1997), according to the formula
\begin{eqnarray}
V_N & = & V_{exp\, N} \,\, (1+\sin i)    \\
V_S & = & V_{exp\, S} \,\, (1-\sin i)    
\label{F-dustv}
\end{eqnarray}
where $V_N$ and $V_S$ are the redshifts of the light scattered from 
the blobs as observed in a specific spectral line (e.g. in the strong
\ha\ line produced in the core), and $i$ is the inclination between
the plane of the sky and the direction of the blobs' motion, that
we assume to be the same for both blobs and coincident with the long
axis of the whole nebula.

Our new deep long-slit spectra allow us to accurately measure $V_N$
and $V_S$ from both the \ha\ and \hb\ lines. Lines profiles in the
core and the blobs are shown in Figure~\ref{F-lineprofile}, and the
redshifts measured by Gaussian fitting are listed in
Tab.~\ref{T-blovel}. As is well known, the \ha\ profile in the core of
\m29\ is complex, showing two peaks and extended wings
(\cite{b89}). This is clearly visible (Figure~\ref{F-lineprofile},
top) at the resolution of our spectra ($\sim$30~\kms). For \hb, the
two components are less well-defined, and the line is
blue-shifted by 15~\kms\ with respect to \ha\ in both the core and the
lobes (Figure~\ref{F-lineprofile}, bottom). This property was
investigated by \cite{t05}, who showed that the trend continues for the
higher-level lines of the Balmer series. The effect was modelled by
means of 3D radiative transfer in a dense disc with expanding wind
velocities slowly increasing with distance from the centre.

\begin{table}[!ht]
\centering
\caption{Velocities of \ha\ and \hb\ profiles in the core and outer
  blobs of \m29\ obtained by Gaussian fit. Line widths (full width at
  a half maximum, corrected for instrumental broadening) are indicated
  in parenthesis. All figures are in \kms\ and heliocentric, except
  for $V_N$ and $V_S$ which are relative to the core profile (see
  text). The symbol ``:'' indicates the uncertain measurement due to
  poor signal in the \hb\ line in the northern blob.}
\begin{tabular}{lrr}
\hline\hline
Structure          & V \ha\  &  V \hb\   \\   
\hline\\[-7pt]                                                    
Core 1$^{st} peak$ &  18 (40) &  10 (40) \\  
Core 2$^{nd} peak$ &  92 (45) &  77 (45) \\
North blob         & 267 (85) & 253:(80) \\
South blob         & 177 (85) & 162 (80)  \\[5pt]
$V_N$              & 192      & 188:      \\
$V_S$              & 102      & 102       \\
\hline
\end{tabular}
\label{T-blovel}
\end{table}

For the present discussion, we note the following. First, line
profiles measured in the blobs do not have the double-peaked shape
shown in the core. This implies that either there is a velocity
gradient within the blobs that smooths out the profile emerging from
the core, or dust in the blobs sees a different core line profile than we
do. For the former hypothesis to be true, we estimate that a velocity
dispersion of a few tens of \kms\ within each blob is needed to dilute
the two peaks in the blobs' profile. This however implies that the
blobs would disperse significantly (over $\sim$10~arcsec for an
internal velocity dispersion of 20~\kms) over their lifetime, unless
they are subjected to some constraining mechanism.
The latter hypothesis is very appealing, as it would mean that a
privileged line of sight from the poles of the nebula with a clear
view of the inner source is available from the light reflected by
the blobs (as in a dentist's tilted mirror). Along this direction, the line
profile might be simpler owing to a smaller optical thickness of the
central disc as seen from the poles.  Radiative transfer modelling
testing this hypothesis would be valuable. Latitude-dependent Balmer
line profiles are also observed in the Homunculus reflection nebula
around $\eta$~Car (\cite{sm03}).


In any case, that both \ha\ and \hb\ in the core have complex
profiles, differently redshifted and morphologically different from
those in the blobs, complicates the estimate of the actual redshift of
the blobs. For example, subtracting the systemic velocity of the
nebula from the observed redshift of each knot could in principle give
a wrong estimate of $V_N$ and $V_S$ in Eqs. (1) and (2). The relevant
figure here is the difference in redshift between the core and the
blobs profile, which is the direct result of the blob motions. We
computed it by shifting the profiles of the blobs in wavelength until
the closest match with the core profile was obtained.  The values of
$V_N$ and $V_S$ computed in this way are also listed in
Tab.~\ref{T-blovel}. We note the excellent agreement between the figures
obtained for \ha\ and \hb. They are also in good agreement with the
values computed by \cite{s97} and \cite{s00}, although they should be
considered more precise given the quality of the data and the more
accurate method adopted.

Assuming that the two blobs are moving away from the central star at
the same speed, i.e. that $V_{exp\, N}=V_{exp\, S}$, an application of
Eqs. (1) and (2) yields an inclination $i=18$\deg\ and a speed of the
blobs $V_{exp}=147$~\kms.  This inclination agrees with the value
$i=17$\deg\ found in the kinematical study of \cite{s00}, which is
only slightly larger than the value $i=15$\deg\ found by \cite{s97},
and more significantly larger than the former determination of
$i=11$\deg\ by \cite{g91}. 
The inclination of \m29\ was also estimated
by fitting inclined circular rings to the parallel lanes visible at
intermediate latitudes in the lobes of \m29\ in the HST \oi\ images.
This favours an inclination of between 19 and 23 degrees.

However, the analysis of the proper motions seems to indicate that the
southern blob is moving twice as fast as the northern blob. If we assume
that the ratio of $V_{exp\, S}$ to $V_{exp\, N}$ is the same as
computed from the proper motion analysis (i.e. 1.8), then the inclination
resulting from applying Eqs. (1) and (2) with this constraint is
as high as 33\deg, with $V_{exp\, N}=124$~\kms\ and $V_{exp\,
  S}=224$~\kms. Such a high inclination can be excluded for the large
disagreement with all other determinations mentioned earlier. We
therefore conclude that the apparent difference in the proper motions
between the north and south blob does not reflect a real difference
in their speed.  We have already mentioned that the difference is,
conservatively speaking, within the measurement errors. In addition,
at least part of the difference -- if true -- could be explained by
internal motions within the blobs. Another effect that may play a role
is that, while Doppler shifts are sensitive to the bulk
motion of the blobs, proper motions can reflect pattern motions, as
for instance in a shock front associated with these supersonic flows.

\subsection{New distance, size, age, and luminosity}

Given the considerations at the end of the previous section, to
compute the distance of \m29\ we combined the determination of the
true speed of the blobs, 147~\kms, their inclination,
18\deg, with the global separation rate of the blobs in the plane of
the sky, $0.046\pm0.004$ arcsec~yr$^{-1}$ for their total separation,
neglecting possible differences between the two blobs. This gives a
distance of 1.3$\pm$0.2~kpc.  The error includes the possibility that
the south blob is indeed moving away from the core at a larger
velocity than the north one, which would also imply that the
inclination of the nebula is larger than the 18\deg\ adopted above
(see discussion in the previous section). Our determination is twice
the value quoted by \cite{s97}. While reviewing their computation, we
detected an error of a factor of two in the calculation; the
additional differences are due to the new proper motion
determination.

With a new distance of between 1.1 and 1.5 kpc, the interblob
separation would be 0.8~pc, and the kinematical age of the blobs
2500~years.  As noted by \cite{sm05}, the blobs would have the same
kinematical age as the inner lobes and the central molecular torus
(Solf 2000). Therefore, all observed morphological components might
have been produced in the same event.

The bolometric luminosity of the system, using the spectral energy
distribution as in \cite{s97}, would be constrained to be between 1600
and 3000~L$_\odot$.  This is consistent with the value of
2500~L$_\odot$ in Lykou et al. (2011), and reopens the possibility --
compared to the low luminosity estimated in \cite{s97} -- that the
central source in \m29\ hosts an AGB or luminous post-AGB star.

\begin{figure}[!t]
\centering
\includegraphics[width=9cm]{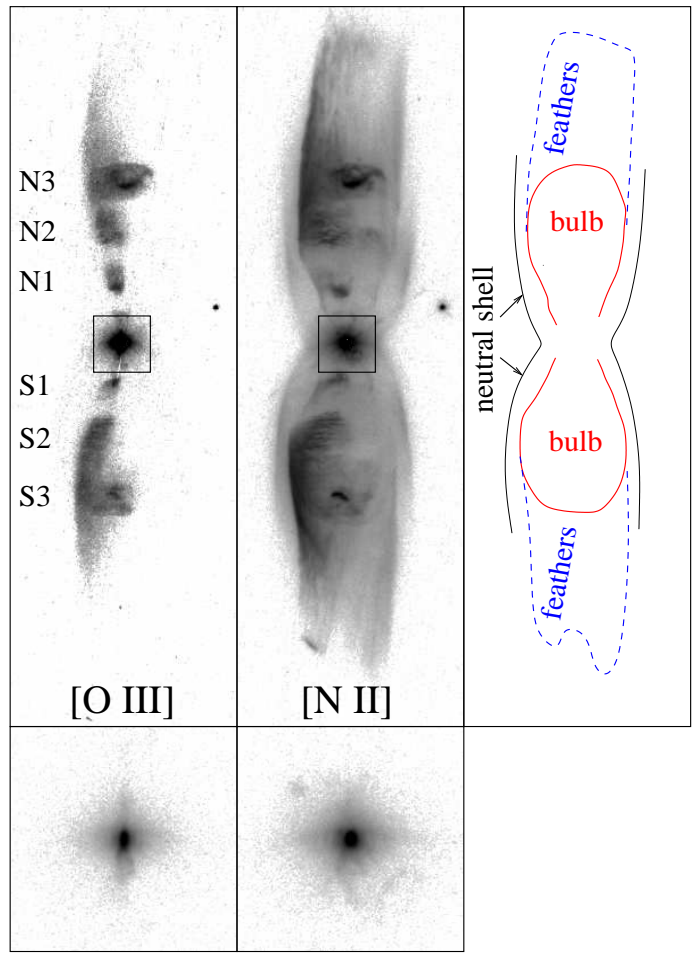}
\caption{Top: HST \oiii\ (left) and \nii\ (centre) images obtained on
  June 2001, in a logarithmic intensity scale. North is up and east to
  the left. The f.o.v. of each box is 20$''$$\times$64$''$. At right,
  a cartoon illustrating the main morphological features of the inner nebula 
  discussed in
  the text. Bottom: 5$''$ $\times$ 5$''$ blowup of the inner regions
  (corresponding to the inset boxes in the upper panels), from the
  dithered HST images obtained on September 2001.}
\label{F-hstima}
\end{figure}

\begin{figure*}[!ht]
\centering
\includegraphics[width=18.5cm]{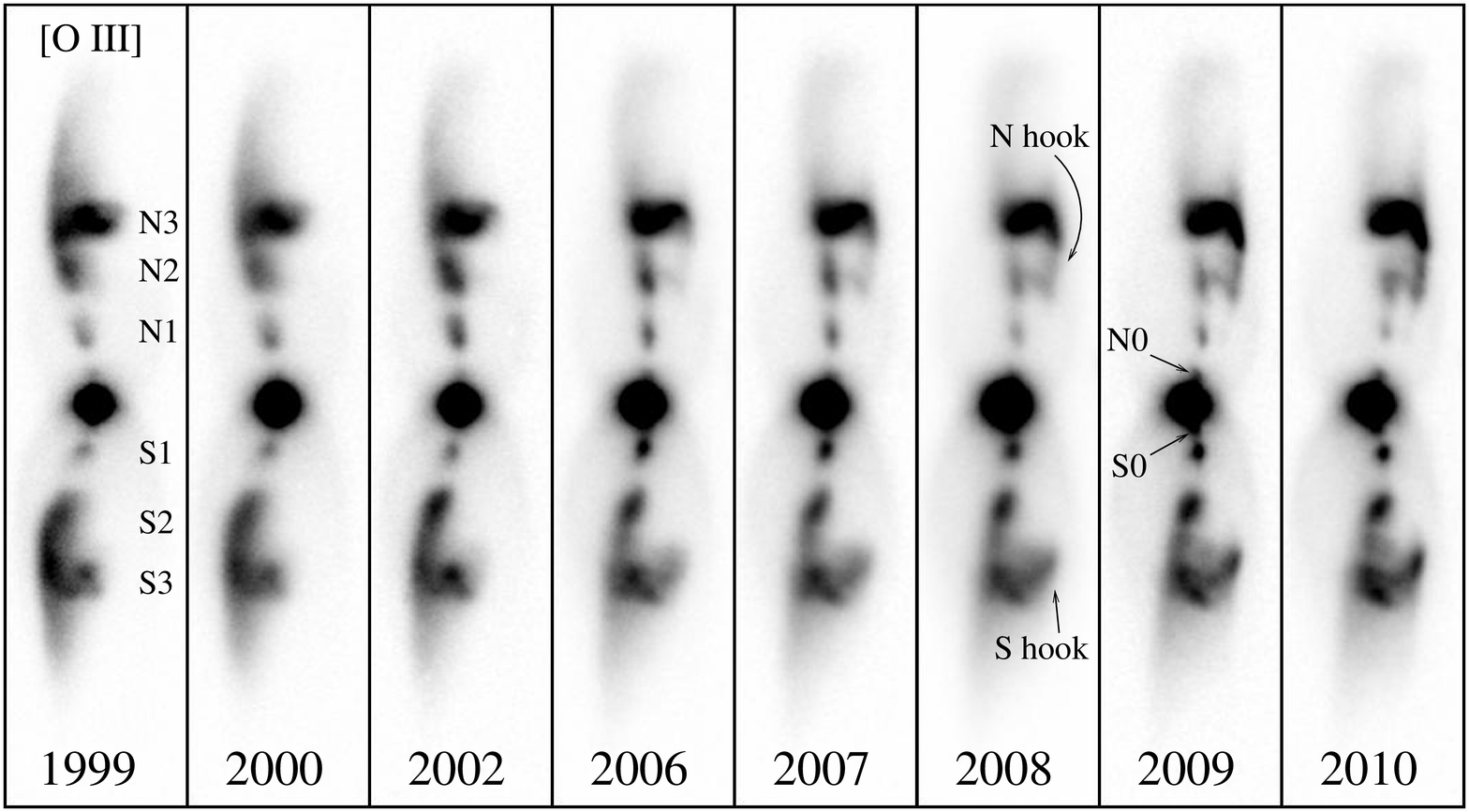}
\includegraphics[width=18.5cm]{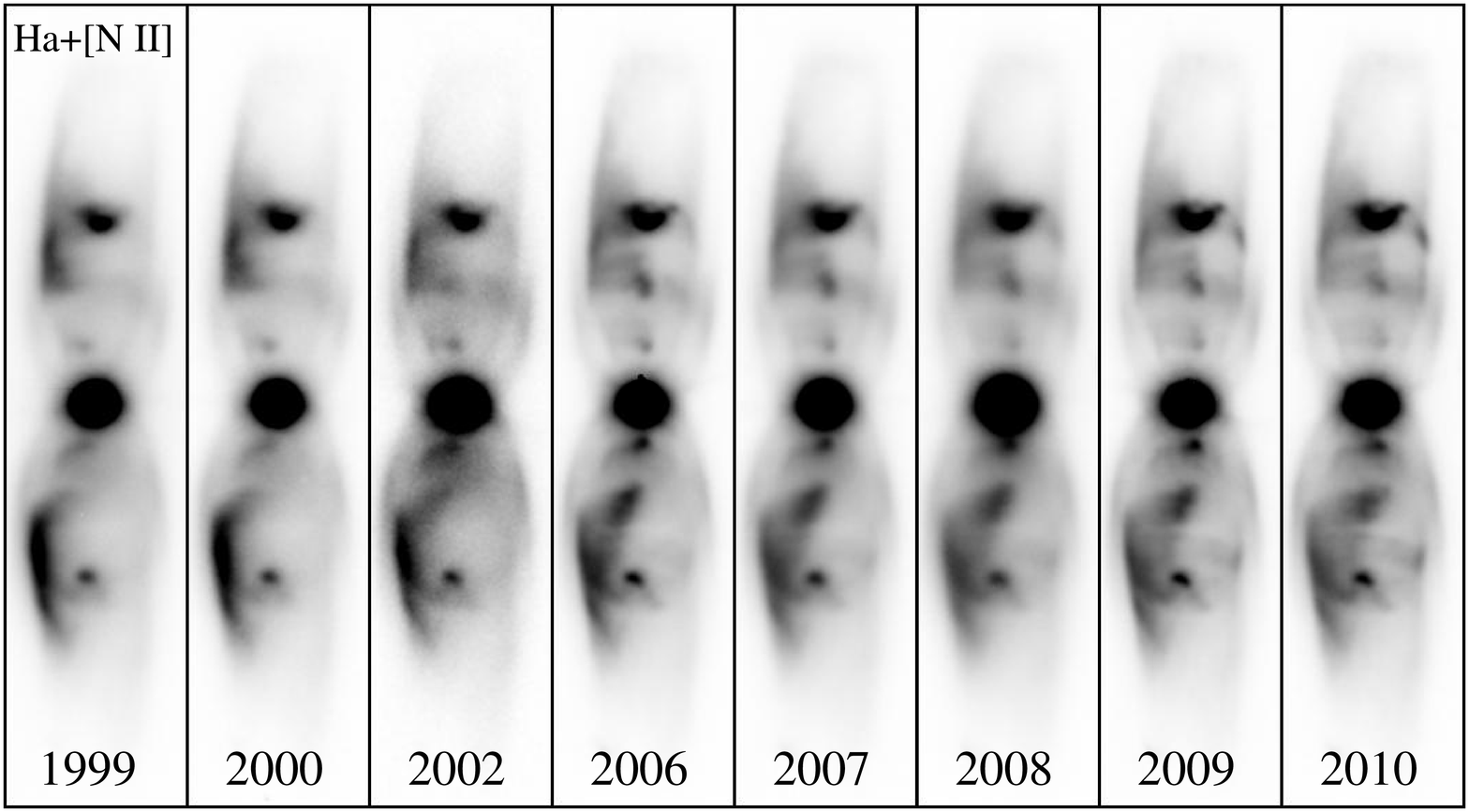}
\caption{{\it (a)} Evolution of the rotating pattern of \m29\ from
  1999 to 2010 in the images from the NOT telescope. \oiii\ and
  \ha+\nii\ images are shown in the upper and lower rows,
  respectively. All images are displayed in a linear intensity
  scale. North is up and east to the left. The f.o.v. of each box is
  14$''$.5$\times$64$''$.\newline {\it (b)} Colour animation obtained
  combining images from 1997 to 2010. 
  Green is the \oiii\ emission, red is
  \ha+\nii. This figure is only available in the electronic version of
  the paper, or at 
\url{http://www.iac.es/galeria/rcorradi/m29}.}
\label{F-sequence}
\end{figure*}

\section{The rotating light pattern}

\subsection{Morphological properties of the features}

The HST images in Figure~\ref{F-hstima}, obtained in 2001, reveal
details of the individual features that move inside the inner lobes of
\m29.  While N1, S1, N3, and S3 appear as true knots that retain their
shapes -- but not their locations -- over time, before 2002 N2 and S2 were 
more clearly described as ``arcs''.  Within N1 and S1, the \nii\ emission lies
closer to the star than the \oiii\ emission (Doyle et al. 2000).  All
these features appear to be located on the walls of cavities within
the inner lobes that are reminiscent of ``bulbs''. These bulbs,
sketched in the right side panel of Figure~\ref{F-hstima} and partly
recognizable in both the HST \nii\ image and the \ha+\nii\ ground-based images 
of Figure~\ref{F-sequence}, stand out
better in the higher resolution \ha\ HST images in Figure~2 of
\cite{d00}, or in the near-IR image obtained by \cite{sm05} in the
\feii~16435~\AA\ emission. At the position of the base of N2 and S2,
filamentary structures with orthogonal extensions similar to
``feathers'' appear to depart from the walls of the bulbs along the
polar directions.

All structures mentioned so far appear in emission lines from ionized
gas. They perhaps represent the ionized inner skin of a thick, 
(mostly) neutral shell, also indicated in the sketch of
Figure~\ref{F-hstima}, which shines in dust-scattered light as well as
in \oi\ and \hh\ emission (Goodrich 1991, Balick 1999, Smith, Balick
\& Gehrz 2005), and forms the outer lobes extending as far as the
outermost dusty blobs discussed in Sect.~\ref{S-blobs}. This
interpretation of the apparently nested systems of lobes of \m29\ as
the result of an excitation gradient was proposed by \cite{sm05}, and
is consistent with the lack of evidence that the various components have 
different kinematical ages.

In the innermost few arcseconds from the central star (bottom panels
of Figure~\ref{F-hstima}), the \nii\ and \oiii\ emission show a faint,
diffuse east-west structure that is aligned with infrared dust
emission (\cite{l11}).  In the vertical direction,
the structure is thinnest in \oiii, and near the centre is not
resolved at the HST resolution (0$''$.1, or 130 A.U. at the adopted
distance). It then seems to open up on both sides with increasing
distance from the central star. In the \nii\ light, the emission is
broader.  The vertical structure points
towards the knots N1 and S1 that were observed at the same time.  

\subsection{Analysis of the rotation}
\label{sS-rotation}

The NOT images from 1999 to 2010, taken with the same instrument and
under similarly good seeing conditions, provide a homogeneous set to
analyse the evolution of the rotating pattern in the inner lobes of
\m29.  The sequence of \oiii\ and \ha+\nii\ NOT images from 1999 to
2010 is shown in Figure~\ref{F-sequence}a. A colour animation of the
evolution combining images from 1997 to 2010 is available in the
electronic version of the paper as
Figure~\ref{F-sequence}b.\footnote{The comparison between the 1997 and 2001 
HST images is also available at 
http://www.astro.washington.edu/users/balick/M2-9/m29\_97\_01\_movie.gif}

The main characteristics of the rotating pattern are clearly visible
in the new sequence of images. The situation was particularly
favourable recently, as the rotating features crossed the inner lobes
from east to west.  With this orientation, the position of the
features on the walls of the bulbs is most accurately determined, as
is their apparent motion in the plane of the sky, which is the fastest
in their ``century orbit''.  The curvature of the knot trajectories
as well as the kinematical study of \cite{s00} indicate that the
motion has been on the side of the bulbs facing the Earth.

We confirm the fundamental properties of the phenomenon described
in \cite{d00}. N3 and S3 are stationary knots at approximately the tip
of the bulbs, but the new \oiii\ images show that their detailed 
morphology changes as the rotating pattern advances westward.  The
arcs N2/S2 and the knots N1/S1 have instead significant lateral motions
(i.e. along latitude circles) on the walls of the bulbs.
Apart from the motions of the dusty blobs N4 and S4, no evidence of 
radial motions (i.e. change of distance from the central star) is 
found for the inner knots (N1 to N3 and S1 to S3) at the
spatial resolution of the ground-based images.

\begin{figure*}[!ht]
\centering
\includegraphics[width=15cm]{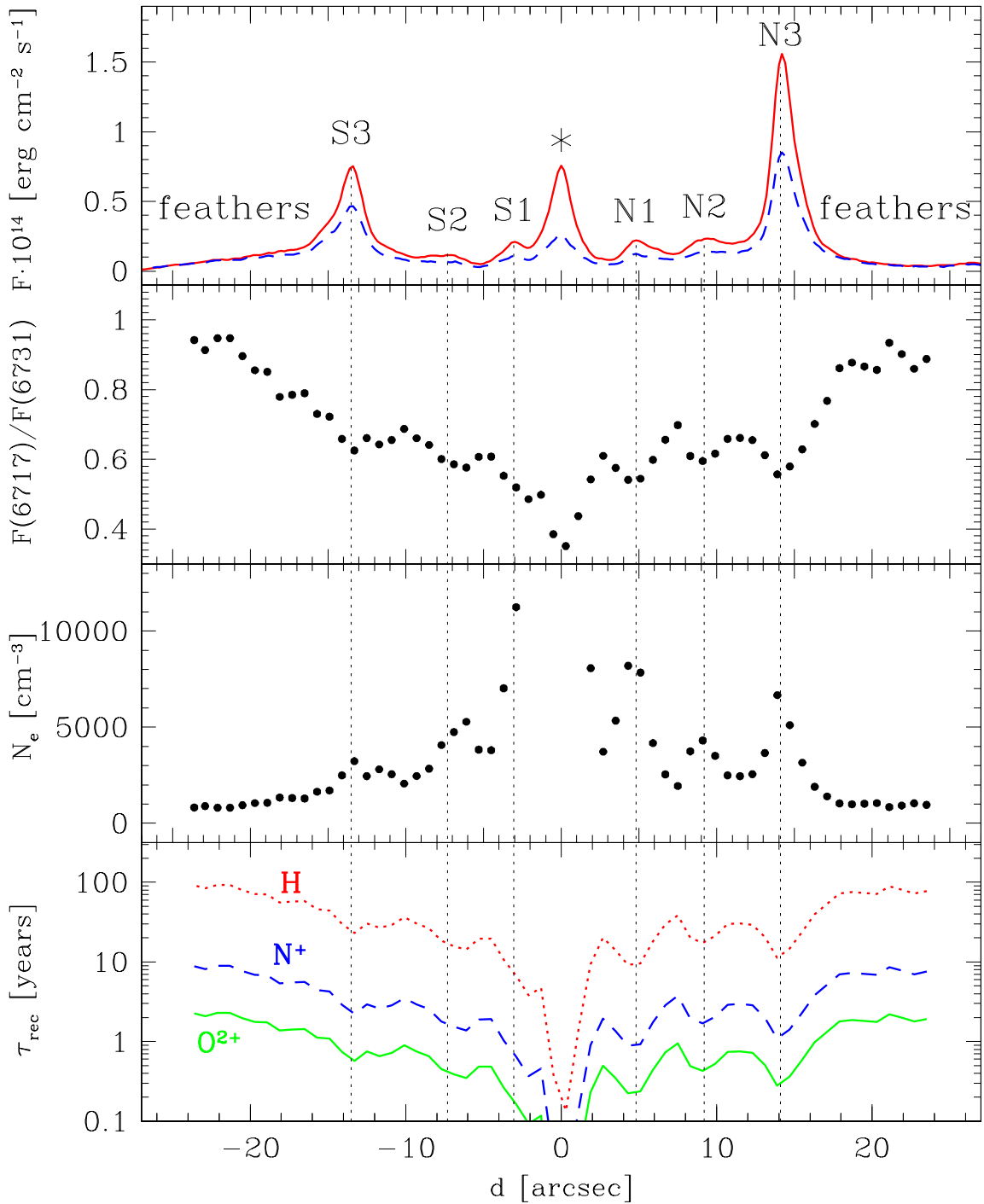}
\caption{Top: The spatial profile of the \sii\ lines emission in the
  2006 WHT spectra. The solid line is the \sii6731 line, and the
  dashed one the 6717 line. Second panel: the F(6717)/F(6731) line
  ratio. Third panel: the electron density $N_e$ profile. Bottom row:
  the corresponding recombination timescales for H, N$^+$, and
  O$^{2+}$.}
\label{F-densities}
\end{figure*}

From Figure~\ref{F-sequence}, it is clear that the \oiii\ emission
provides clearer insight into these motions. This can be explained by different
recombination timescales for different ions, assuming that the gas on
the walls of the bulbs is excited by a rotating ionizing beam 
emitted by the central source.  The spatial
profile of the \sii6717 and \sii6731 lines along the slit of the 2006
WHT spectra, the 6717/6731 line flux ratio, and the corresponding
electron density $N_e$ profile through the main moving features of
\m29\ are shown in the first three rows of
Figure~\ref{F-densities}. In the bottom panel, the recombination
timescales $\tau_{rec}$ are plotted for the H, N$^+$, and O$^{2+}$ ions at the
computed $N_e$ and for an electron temperature of the inner lobes of
10\,000~K (\cite{tr95}).  $N_e$ is about $10^4$~cm$^{-3}$ at the
positions of N1/S1, $5\cdot 10^3$~cm$^{-3}$ at the base of S2/N2, and
of the order of $10^3$~cm$^{-3}$ along the feathers.  At a density of
$10^4$~cm$^{-3}$, $\tau_{rec}$ for the O$^{2+}$ ion is
only two months.  This means that the \oiii\ emission switches off as
soon as the ionizing beam sweeps past.  By comparison, 
at these densities $\tau_{rec}$ is nine
months for the N$^{+}$ ion, and seven years for hydrogen.  This
explains the tails of emission seen eastward of the moving features in
the \ha+\nii\ images, and why the tails of S2/N2 (which are half as 
dense, implying twice $\tau_{rec}$) are more
pronounced than those of N1/S1. For the feathers, ten times less dense
than N1/S1, the recombination time in \ha\ is so long that their
\ha\ emission should persist for a century or so.  Thus, the
\oiii\ line is the most faithful marker of the present position of the
exciting beam.

\begin{figure}[!ht]
\centering
\includegraphics[width=9cm]{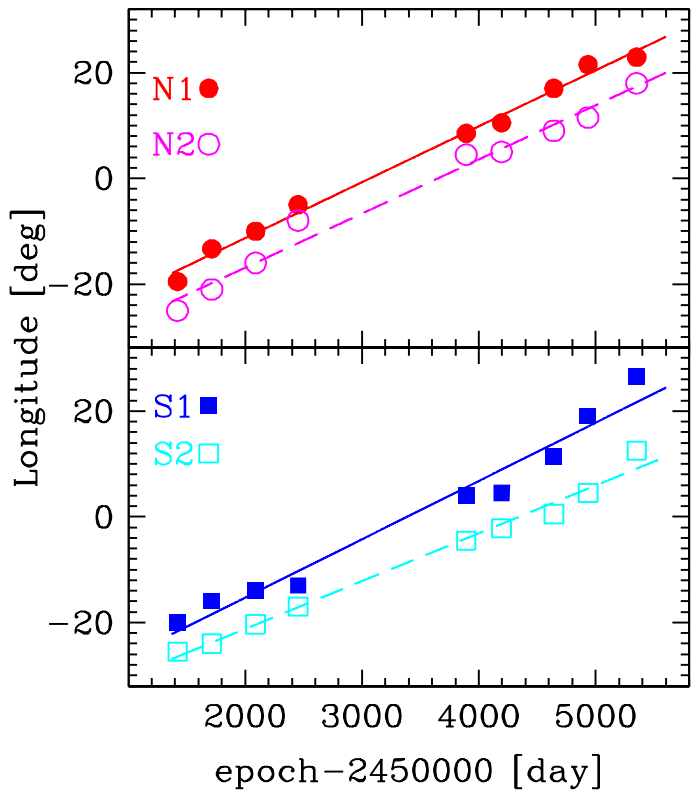}
\caption{Variation in the longitude of the rotating features of
  \m29\ between 1999 and 2010. The error in each longitude
  determination varies with feature and epoch, but on average is of a
  few degrees. The linear fits are also shown as solid and dashed lines.}
\label{F-rotation}
\end{figure}

Taking advantage of the favourable orientation of the rotating
features, we quantified the concept presented by \cite{d00} of a
corkscrew geometry of the pattern. In particular, we focussed our
attention on features N1 and S1, and on the base of features N2 and
S2, whose position is the most accurately measured. Outside N2 and S2, 
the pattern becomes more diffuse. We assume axial symmetry for the
bulbs and the inclination determined in Sect.~\ref{S-blobs}. The
contour of the bulbs is drawn taking advantage of the high resolution
HST images. Under these assumptions, any feature can be located
unambiguously on the walls of the bulbs, and its latitude and
longitude determined.  The longitude, which is the crucial descriptive
parameter, was determined at each epoch by the Gaussian fitting of the
position in the \oiii\ image of the base of features N1, N2, S1, and
S2. Longitude zero is set at ``inferior conjunction'' of the features,
i.e. at their closest passage to the Earth through the plane that is
perpendicular to the plane of the sky and contains the symmetry axis
of the bulbs.
Uncertainties mainly come from the asymmetry of the features.  Images
before 1999 were discarded because the determination of longitudes
is more uncertain because of projection effects.
The variation in longitude with time is shown in
Figure~\ref{F-rotation}.  A delay of N2 with respect to N1, as well as
S2 with respect to S1, is clearly evident. In other words, excitation of the
\oiii\ emission at a given longitude of the bulbs occurs later at the
higher latitudes of N2 and S2 than at the lower latitudes of N1 and
S1.  This confirms the results of \cite{d00}. Averaging the linear fit
of the motion of N1 and S1, which are the sharpest and hence provide
the most reliable measurements, gives a period for the rotating
pattern of 92$\pm$4~years. Passage at longitude zero occurred at the
beginning of 2004 for N1, and at the end of 2005 for S1. The average
delay of N2 and S2 to reach the same longitude as N1 and S1,
respectively, is 1.9~years. The apparent tendency to increase in the
southern side in the past three years needs to be confirmed by future
data.

\begin{figure}[!ht]
\centering
\includegraphics[width=9cm]{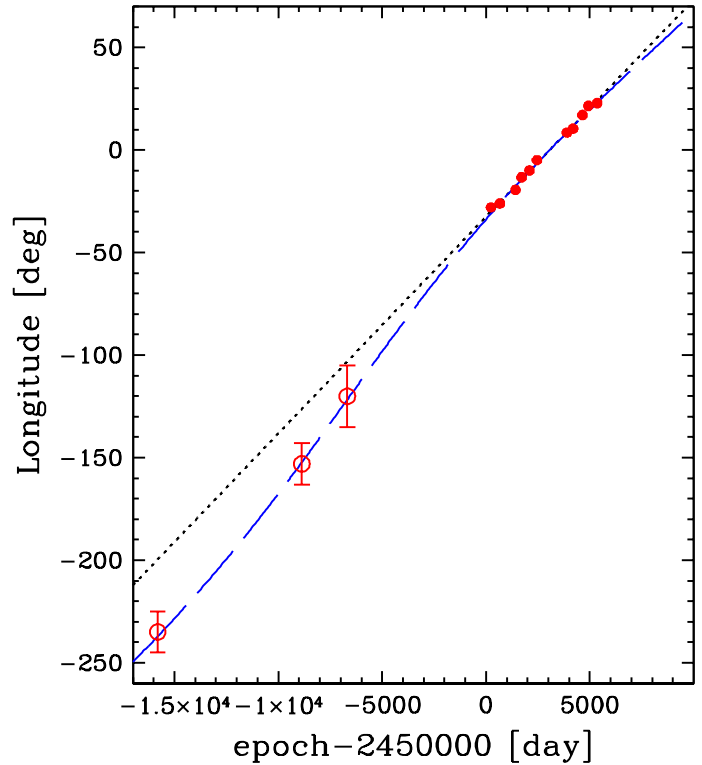}
\caption{Longitude motion of knot N1, including all our data and those
obtained from the analysis of the photographic images of 1952,
  1971, and 1977 (empty circles with errorbars). The dot line is
  the linear fit obtained using the CCD images from 1999 and 2010 as in
  Figure\ref{F-rotation}. The long-dash (blue) line is 
the fit for an eccentric orbit of period 86~yr, eccentricity 0.1, and 
passage at periastron in 1977.}
\label{F-rotation-old}
\end{figure}

The older photographic images from 1952, 1971, and 1977 in \cite{as72}
and \cite{ks80} can be used to check the rotational period determined
above.  Given the limited quality of those images, both intrinsic and
related to the reproduction on paper in the articles, a useful
determination of the longitude can be attempted only for knot N1.
Figure~\ref{F-rotation-old} shows the estimated position of N1 on the
older dates (empty circles with errorbars), compared to the fit of the
motion (dotted line) determined using the 1999-2010 measurements
(dots) as in Figure~\ref{F-rotation}. It appears that, even
considering the relatively large errors (10 to 15 degrees), the old
measurements lie below the fitting line, indicating a steeper slope
than presently measured, and therefore suggesting that the rotation of
the pattern was faster in the preceding $\sim$50~years than in the
period 1999-2010. According to the hypothesis that the rotating pattern is
directly linked to the orbital motion in a binary system, as discussed
in the next section, these angular speed variations may indicate that
the orbit is eccentric. With only three rather uncertain points
available before 1996, the orbital parameters cannot be tightly
constrained. Even so, a small eccentricity $e=0.10\pm0.05$
(corresponding to an axial ratio of 0.995 for the orbit ellipse) is
enough to reproduce the data, as shown by the long-dashed blue line in
Figure~\ref{F-rotation-old}. The orbital period would then be
$P=86\pm5$~years, and the passage at periastron would have occurred
between 1968 and 1984.

A remarkable and unexpected pair of new features resembling ``fishing
hooks'' appeared after 2005. They consist of arcs departing from S3
and - more clearly - from N3, extending along the west edge of the
bulbs.  They are most clearly seen in \oiii\ in
Figure~\ref{F-sequence} (labelled ``N hook'' and ``S hook''), and have
become progressively brighter spanning an increasing range
of latitudes. On the northern side, the feature has developed a bridge
to N2. While the fishing hooks S1 and S2 have progressively
brightened, knots N1 and N2 seem to have faded considerably from 2006
on.  In addition, a pair of inner knots have appeared in the past
three years in the \oiii\ images at the edge of the nuclear PSF. These
are not apparent in the 2001 HST image of Figure~\ref{F-hstima}, and
are labelled as N0 and S0 in Figure~\ref{F-sequence}.

\section{Discussion: rotating jet or light beam?}

As discussed in \cite{d00}, the delay between N1 and S1 (and S1 and
S2) is a measure of the travel time (thus of the speed) of the
ionizing beam, if the beam is ejected
meridionally. With the distance and inclination determined in
Sect.~\ref{S-blobs}, this would result in a remarkable beam speed
between 11\,000 and 16\,000~\kms, supporting the conclusion of
\cite{d00} that the ionizing beam is a collimated spray of high
velocity particles emitted by the central source of \m29. In other
words, a highly supersonic, tenuous, and rotating jet would shock and
excite the walls of the inner cavity of \m29\ (the bulbs) as it hits
them.  For the alternative hypothesis of a beam of energetic {\it
  photons}, the travelling time from N1 to N2, or S1 to S2, would
instead be of only one month, rather than the 1.9~years measured.

In this scenario, the walls of the inner cavity are ionized by the
shocks produced by the impact of the jet particles.  The possibility
that the cavity is excavated (inflated) by the jet itself should also
be considered (\cite{s02, s04}). This would explain the formation of
the inner lobes and the rotating pattern.  If the outer lobes are the
result of an excitation gradient and not independent structures
(sect.~4.1, Smith, Balick \& Gehrz 2005), then no additional
mechanisms other than jet shaping would be needed to explain the main
morphological structures of \m29.

The resulting nebular spectrum would depend on several parameters,
among others the jet properties, the physical conditions of gas on the
walls of the cavity, and the geometry that determines the angle of
incidence of the jet onto the walls. In this respect, N1, S1, N2, and
S2 seem to be located at latitudes corresponding to inflexion points
in the curvature of the bulbs (see the \nii\ image in
Figure~\ref{F-hstima}). In addition, the ``empty'' regions in the rotating
pattern between the core and N1, between N1 and N2 (and the
symmetrical ones in the southern side) correspond to slopes of the
walls almost radially directed from the central star. It is possible
that the appearance/disappearance of the rotating pattern with
latitude (and its ionization structure, with \nii\ closer to the star
than \oiii\ in S1 and N1) is related to different incidence angles
(high/low, respectively) of the jet on the walls of the cavity,
i.e. different entrance velocity of the jet into the upstream
material.

We can imagine the following process for the formation of knots N3 and
S3 and the ``fishing hooks'', which seem confined within the bulbs.
The prominent \feii1.64~$\mu$m emission most likely indicates the
existence of shocks all around the edge of the bulbs. The dense layer
formed in the shocks prevents any ionized gas created by the jet
escaping. Therefore, some of the created double-ionized oxygen might flow 
to higher latitudes along the cavity interior, after receiving a
little momentum from the high-speed particle beam. If the density of
this gas drops a bit, it recombines relatively slowly and persists for
a longer time than the knots N1, N2, S1, and S2. It would therefore drift
toward the polar caps of the bulbs becoming visible as N3 and S3, and
beyond them on the opposite wall (which presently corresponds to the
western side) giving rise to the ``fishing hooks''. In other words,
the bulb walls would force hot gas to slide along them, and form
N3, S3, and the ``fishing hooks''. The bridge to N2 could also be hot
gas that it is sprayed after the beam shocks N2.
The formation of the feathers, which appear to extend outside the
bulbs along the inner skin of the extended lobes of \m29, is more
difficult to imagine: fully 3-D radiative-hydrodynamical modelling is
needed to explain them and more in general the complexity of the
phenomena observed. 

In the proposed scenario, it is possible that no significant stellar
ionizing radiation is emitted or escapes from the dense cocoon of gas
that surrounds the star.
Shocks from the jet produce most of the ionization observed, which at the
density of the gas results in short-lived \oiii\ emission (due to its
fast recombination time) and more long-lived \ha\ emission, persisting
enough that hydrogen never becomes entirely neutral before the jet
swings around again in its ninety-year rotation. Even if soft stellar
UV radiation helps in preventing the singly-ionized gas from fully
recombining, cooling, and becoming neutral (\cite{tr95}), its role
would not be predominant. 
In a future paper, we will present a detailed spectroscopic analysis
of the ionized gas in the rotating pattern, in order to test -- among
other things -- the hypothesis that shock excitation by a jet is
primarily responsible for the observed emission in the lobes.

Another basic property of the rotating pattern of \m29\ is its mirror
symmetry with respect to the the equatorial plane of the nebula. A
two-sided jet in which each side bends toward the same direction as it
travels through a circumstellar medium with a strong density gradient
(produced for instance by the massive wind of a red-giant companion)
can explain this property (\cite{ls01, sb06}). This would directly
link the observed motions to the orbital period of a symbiotic-like
binary central star. If so, the difference in the angular velocity of
the pattern over the past ten years and the period covered by the
oldest images (1950-1980) may indicate that the orbit has a modest
eccentricity.

The alternative scenario is the one proposed by several authors
(e.g. \cite{ls01}), in which the rotating pattern is
photoionized. In this case, a mirror-symmetric rotating jet is also 
invoked to clear a path for the ionizing radiation from the central
star in the dense circumstellar environment along the directions where
the \oiii\ emission is observed.  However, for the time being we
favour the idea that the jet particles themselves rather than photons
ionize the nebula, because in this way the delay of the propagation of
the pattern with latitude is a natural consequence of the limited
speed of the jet compared to light. In the case of a light beam, a
complex system of ``holes'' in the circumstellar cocoon (with a
geometry that does not change with the rotation of the whole
structure) would be needed to photoionize simultaneously the knots
observed at different latitudes and longitudes.

If the cause of the phenomenon is a rotating jet, its $\ge$10\,000
\kms\ speed would be unusually high for planetary nebulae and
pre-planetary nebulae, where the observed highly collimated outflows
are one or two orders of magnitude slower.  Jets in symbiotic stars
have generally larger velocities up to several thousand
\kms\ (\cite{sok04}), but still lower than the figure that we estimate
for \m29.  This makes \m29\ even more unusual, and the discussion of
its nature more complex. The $\sim$90-yr period of the rotating
pattern, the presence of a jet, the nuclear spectrum, and the recent
discovery of circumstellar discs (\cite{l11}), suggest the presence of
a symbiotic binary star at its centre: a compact star, perhaps a
relatively cool white dwarf, accreting from the wind of a giant and
producing the rotating jet. System parameters for our revised distance
of 1.3~kpc, such as the bolometric luminosity or the circumbinary disc
properties in the models by Lykou et al. (2011), are consistent with
this hypothesis.

While IR observations further constrain the possible parameters of the
binary system (e.g. the stellar masses, \cite{l11}), the conclusive
evidence of binarity, i.e. the detection of a luminous red giant
companion, is still missing. Spectroscopy and photometric monitoring
at all wavelengths have failed to detect either the spectral features
or variability typical of a cool giant (\cite{san09}). The
characteristics of the accreting component are poorly known too. New
efforts to unveil the central stars should be considered as a priority
for the future.

\begin{acknowledgements}

We are very grateful to the staff of the Nordic Optical Telescope, and
in particular Thomas Augusteijn and Johannes Andersen, for making the
long-term image monitoring of \m29\ possible through the NOT
fast-tracking programme and discretionary time.  We thank the
referee, Dr. Olivier Chesneau, for his valuable suggestions.

The work of RLMC and MSG is supported by the Spanish Ministry of
Science and Innovation (MICINN) under the grant AYA2007-66804.
Support for program 9050 was provided by NASA through a grant from the
Space Telescope Science Institute, which is operated by the
Association of Universities for Research in Astronomy, Inc., under
NASA contract NAS 5-26555.
\end{acknowledgements}


\begin{thebibliography}{} 

\bibitem[Acker et al. 1992]{a92}
Acker, A., Marcout, J., Ochsenbein, F., Stenholm, B., Tylenda, R. 
1992, Strasbourg-ESO catalogue of galactic planetary nebulae,
ESO Garching
\bibitem[Aller and Swings (1972)]{as72}
Aller, D.A., Swings, J.P. 1972, ApJ, 174, 583
\bibitem[Balick 1989]{b89}
Balick, B., 1989, AJ, 97, 476
\bibitem[Balick \& Hajian 2004]{bh04}
Balick, B., Hajian, A.R. 2004, AJ, 127, 2269
\bibitem[Chesneau et al. 2007]{ch07}
Chesneau, O., Lykou, F., Balick, B., et al. 2007, A\&A, 473, L29
\bibitem[Cohen et al. 2011]{co11}
Cohen, M., Parker, Q.A., Green, A.J., et al. 2011, MNRAS, in press
\bibitem[Doyle et al. (2000)]{d00} 
Doyle, S., Balick, B., Corradi, R.L.M., Schwarz, H.E. 2000, AJ, 119, 1339
\bibitem[Goodrich (1991)]{g91} 
Goodrich, R.W. 1991, ApJ, 366, 163
\bibitem[Kohoutek \& Surdej (1980)]{ks80}
Kohoutek, L., Surdej, J. 1980 A\&A, 85, 161
\bibitem[Lykou et al. 2011]{l11}
Lykou, F., Chesneau, O., Zijlstra, A.A., 
et al. 2011, A\&A, in press
\bibitem[Livio and Soker 2001]{ls01}
Livio, M., Soker, N. 2001, ApJ, 552, 685
\bibitem[Monet et al. 2003]{m03}
Monet, D.G., Levine, S.E., Canzian, B., et al. 2003, AJ 125, 984
\bibitem[Santander-Garc\'\i a, Corradi \& Mampaso 2009]{san09}
Santander-Garc\'\i a, M., Corradi, R.L.M., Mampaso, A. 2009, in
``Asymmetrical Planetary Nebulae IV'', Corradi, R.L.M., Manchado,
A. \& Soker, N. eds., IAC, p. 555 ({\it
  http://www.iac.es/proyecto/apn4/pages/proceedings.php})
\bibitem[Schmeja \& Kimeswerner (2001)]{sk01}
Schmeja, S., Kimeswerner, S. 2001, A\&A, 377, L18
\bibitem[Schwarz et al. (1997)]{s97}
Schwarz, H.E., Aspin, C., Corradi, R.L.M., Reipurth, B. 1997, A\&A, 319, 267
\bibitem[Smith et al. 2003]{sm03}
Smith, N., Davidson, K., Gull, T.R., Ishibashi, K., Hillier, D.J. 
2003, ApJ, 586, 432
\bibitem[Smith et al. 2004]{sm04}
Smith, N., Morse, J.A., Collins, N.R., Gull, T.R. 2004, ApJ, 610, L105
\bibitem[Smith, Balick \& Gehrz (2005)]{sm05}
Smith, N., Balick, B., Gehrz, R.D. 2005, AJ, 130, 853
\bibitem[Soker 2002]{s02}
Soker, N. 2002, ApJ, 568, 726
\bibitem[Soker 2004]{s04}
Soker, N. 2004, A\&A, 414, 943
\bibitem[Soker and Bisker 2006]{sb06}
Soker, N., Bisker, G. 2006, MNRAS, 369, 1115
\bibitem[Sokoloski et al. 2004]{sok04}
Sokoloski, J.L., Kenyon, S.J., Brocksopp, C., Kaiser, C.R., Kellogg, E.M. 
2004, RevMexAA (Series de Conferencias), 20, 35
\bibitem[Solf (2000)]{s00}
Solf, J. 2000, A\&A 354, 674
\bibitem[Torres-Peimbert et al. (2005)]{t05} 
Torres-Peimbert, S., Arrieta, A., Georgiev, L., Bautista, M. 2005, in
Planetary Nebulae as Astrophysical Tools, Szczerba R. \& Stasinska
G. eds., American Institute of Physics, p.148
\bibitem[Trammell, Goodrich \& Dinerstein 1995]{tr95}
Trammell, S.A., Goodrich, R.W. Dinerstein, H.L. 1995, ApJ 453, 761 
\bibitem[Zweigle et al. (1997)]{z97}
Zweigle, J., Neri, R., Bachiller, R., Bujarrabal, V., Grewing, M. 
1997, A\&A, 324, 624
\end{thebibliography}
\end{document}